\documentclass[prl,twocolumn,showpacs,floatfix,amsmath,amssymb,superscriptaddress]{revtex4}
\usepackage{graphicx}
\usepackage{amsmath, eufrak}
\usepackage{color}
\usepackage{dcolumn}
\usepackage{bm}
\usepackage{subfigure}

\def\etal{~\textit{et~al}}
\def\hc{{\rm H.c.}}
\def\ra{\rangle}
\def\la{\langle}

\begin{document}

\title{Possible Exciton Bose Liquid in a hard-core boson ring model}

\author{Tiamhock Tay}
\author{Olexei I. Motrunich}

\affiliation{Department of Physics, California Institute of Technology, Pasadena, CA 91125}

\date{\today}


\begin{abstract}
We present a Quantum Monte Carlo study of a hardcore boson model with ring-only exchanges on a square lattice, where a $K_1$ term acts on 1$\times$1 plaquettes and a $K_2$ term acts on 1$\times$2 and 2$\times$1 plaquettes.  At half-filling, the phase diagram reveals charge density wave for small $K_2$, valence bond solid for intermediate $K_2$, and possibly for large $K_2$ the novel Exciton Bose Liquid (EBL) phase first proposed by Paramekanti\etal~[Phys.~Rev.~B {\bf 66}, 054526 (2002)].  Away from half-filling, the EBL phase is present already for intermediate $K_2$ and remains stable for a range of densities below 1/2 before phase separation sets in at lower densities.
\end{abstract}
\maketitle


The search for exotic quantum phases has gained a very wide audience in recent years.   There has been much interest in ``Bose-metal''--type phases in boson and spin systems, and more broadly non-Fermi liquids in electron systems. The interest has also spread to the High Energy community where ``holographic metals'' have been much investigated.   To date, there is still no microscopic model where a Bose-metal phase can be controllably demonstrated.   The novel Exciton Bose Liquid (EBL) proposed a number of years ago by Paramekanti \etal \cite{Paramekanti2002} provides an interesting critical bosonic phase which shares many characteristics normally associated with electrons in a metal \cite{Sachdev2002} and which can be viewed as a Bose-metal in a restricted sense.   Specifically, it is an example of a 2D quantum system with microscopic bosonic degrees of freedom that nevertheless has surfaces of low-energy excitations and metal-like properties such as large specific heat and thermal conductivity.   The striking proposal of Paramekanti\etal \cite{Paramekanti2002} stimulated a number of works seeking to establish the stability of the EBL phase in hard-core boson models with ring exchange interactions on the square lattice \cite{Sandvik2002, Melko2004, Rousseau2004, Rousseau2005}.  However, these studies found that the EBL is not realized in these models.    Instead, such ring interactions favor a $(\pi, \pi)$ charge density wave (CDW) in the half-filled case, while away from half-filling they induce strong tendencies to phase separation.

Here we propose a ring-only hard-core boson model on the square lattice with additional ring exchanges as a candidate model for realizing the EBL.   We define plaquette exchange operators
\begin{equation}
  P^{mn}_{\bf r} = b_{\bf r}^\dagger~ b_{{\bf r}+m{\bf \hat{x}}}~ b_{{\bf r}+m{\bf \hat{x}}+n{\bf \hat{y}}}^\dagger~ b_{{\bf r}+n{\bf \hat{y}}} + \hc ~,
\end{equation}
where $b_{\bf r}$ annihilates a boson on a site ${\bf r}$, and ${\bf \hat{x}}, {\bf \hat{y}}$ are the unit vectors on the square lattice.  The Hamiltonian is
\begin{equation}
  \hat{H} = -K_1\sum_{\bf r} P^{11}_{\bf r} - K_2 \sum_{\bf r}\left(P^{12}_{\bf r} + P^{21}_{\bf r}\right) ~.
\end{equation}
Throughout we set $K_1 = 1$.  The original ring model proposed in Ref.~\onlinecite{Paramekanti2002} and studied numerically in Refs.~\onlinecite{Sandvik2002, Melko2004, Rousseau2004, Rousseau2005} corresponds to $K_2 = 0$.  The extra ring terms frustrate the CDW order preferred by the $K_1$ term and may allow the EBL phase to be realized for large $K_2$ at half-filling.  The present $K_1$-$K_2$ model has the same lattice symmetries and boson number conservation on each row and column as the original model.  From the outset, we define our Hilbert space as the sector with equal number of bosons in each row and column.   Note that we are allowed to make such a sector restriction consistent with the conservation laws of the Hamiltonian, and it is in this setting that we look for a uniform ``liquid'' phase of bosons that does not break any lattice symmetries.  The original reference \cite{Paramekanti2002} envisioned the EBL that would be stable also in the presence of boson hopping.  While we do not think that our $K_1$-$K_2$ model will achieve this, realizing the EBL even in the restricted sense is already very interesting as it is expected to have many unusual properties described earlier, which can then be confronted in a concrete 2D system.

For simulation on large two-dimensional lattices, the Stochastic Series Expansion (SSE) has been the method of choice. Although the model does not have a sign problem, implementation issues have so far precluded the study of ring-only models using SSE \cite{Sandvik2002}. We instead apply the Green's Function Monte Carlo (GFMC) method using the bias-controlled technique of Ref.~\cite{Buonaura1998} for an exploratory smaller scale study.  To obtain an initial wavefunction for the GFMC projection, we first perform a Variational Monte Carlo study using EBL-inspired wavefunctions.  Specifically, consider a quantum rotor model which is a soft-core version of the $K_1$-only model:
\begin{eqnarray}
  \hat{H}_{\rm rotor} &=& -K \sum_{\bf r} \cos\left( \phi_{\bf r} - \phi_{\bf r+\hat{x}} + \phi_{\bf r+\hat{x}+\hat{y}} - \phi_{\bf r+\hat{y}} \right)\nonumber\\
&+& \frac{U}{2} \sum_{\bf r} \left( n_{\bf r} - \bar{n} \right)^2 ~,
\label{Hrotor}
\end{eqnarray}
with canonically conjugate phase $\phi_{\bf r}$ and boson number $n_{\bf r}$ operators.  A harmonic ``spin wave'' theory where we expand the cosines motivates the following wavefunction, which we now restrict to the Hilbert space of the hard-core boson model \cite{Motrunich2007}:
\begin{eqnarray}  
  \Psi_{\rm EBL} &\propto& \exp\left[ -\frac{1}{2}\sum_{\bf r,r'} u({\bf r} - {\bf r'})~n_{\bf r} n_{\bf r'} \right] \label{PsiEBL},\\
  u({\bf r}) &=& \frac{1}{L^2} \sum_{\bf q} \frac{W({\bf q})~ e^{i {\bf q} \cdot {\bf r}}}{4\left\vert \sin(q_x/2) \sin(q_y/2) \right\vert} ~.
\label{eq:EBL_pseudo_potential}
\end{eqnarray}
In the $K_1$-only spin wave theory, $W({\bf q}) = \sqrt{U/K}$.  For the $K_1$-$K_2$ case, $W({\bf q})$ becomes a {\bf q}-dependent function with two variational parameters.  The formal EBL wavefunction is quite interesting by itself --- e.g., it can produce both liquid and solid phases \cite{long_paper}.  Here it is only important that the optimal trial wavefunctions provide good starting points for the GFMC projection.  We tested our GFMC setup against exact diagonalization calculations for a 6$\times$6 lattice and obtained complete agreement for all physical quantities considered in this work.  

To characterize the system, we measure the density structure factor
\begin{equation}
  S(q_x,q_y) = \frac{1}{L^2}\sum_{\bf r,r'} e^{i{\bf q}\cdot({\bf r}-{\bf r'})}\la n_{\bf r}n_{\bf r'}-\bar{n}^2 \ra ~,
\end{equation}
and the plaquette structure factor
\begin{equation}
  P(q_x,q_y) = \frac{1}{L^2}\sum_{\bf r,r'} e^{i{\bf q}\cdot({\bf r}-{\bf r'})}\la (P^{11}_{\bf r})^2 ~(P^{11}_{\bf r'})^2 \ra ~,\label{Eq:placket_SF}
\end{equation}
where $(P^{11}_{\bf r})^2$ equals $1$ if the 1$\times$1  plaquette is hoppable and $0$ otherwise.  
This operator is diagonal in the boson number basis and is easier to implement in the GFMC.  While quantitatively different from the off-diagonal $P^{11}_{\bf r}$ plaquette structure factor used in Ref.~\onlinecite{Sandvik2002}, both give qualitatively the same access to bond-solid--type phases.

\begin{figure}
  \centering
  \includegraphics[trim=0 20 0 0,width=\columnwidth]{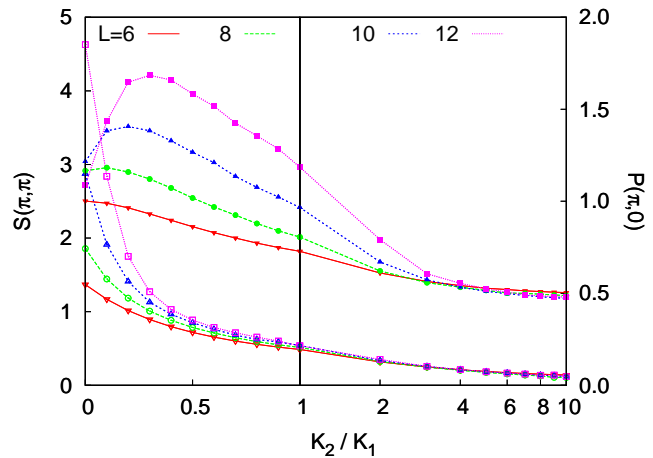}
  \caption{[Color online] Structure factors versus $K_2$ for periodic lattices with length $L=6,8,10$ and $12$, where open symbols are for the density structure factor $S(\pi, \pi)$, and solid symbols are for the plaquette structure factor $P(\pi, 0)$. Note that linear scale is used for $K_2=0$ to $1$ in steps of $0.1$, and log scale for $K_2=1$ to $10$ in steps of $1$.}
  \label{fig:SF_crossover}
\end{figure}

We now summarize our main results at density 1/2.  For $K_2=0$, the GFMC study confirms the staggered CDW phase anticipated by Paramekanti\etal\cite{Paramekanti2002} and observed by Sandvik\etal\cite{Sandvik2002} in the $J$-$K$ model at $K/J=64$, where $J$ is the nearest neighbor hopping strength.  We find that as $K_2$ increases, the ground state changes to a Valence Bond Solid (VBS) at $K_2 \sim 0.4$.  Interestingly, Sandvik\etal\cite{Sandvik2002} observed a striped bond-plaquette order in the $J$-$K$ model for $8 \lesssim K/J \lesssim 14$.  (The two findings may be connected if, in a perturbative picture in the insulator, $J$ induces significant $K_2$ ring exchanges.)  As $K_2$ increases further, our GFMC study suggests that the bond order vanishes for $K_2 \gtrsim 4$, thus allowing the possibility for a realization of the EBL phase at half-filling.

On the other hand, for a window of density below 1/2, we already find a clear EBL phase for a wide $K_2$ range including also intermediate $K_2$ values. At still lower density the system has strong tendencies to phase-separate, which occurs since ring exchanges are operative only when bosons are sufficiently close \cite{Rousseau2004, Rousseau2005}.  Below, we present evidence for these conclusions.

\begin{figure}
  \centering
  \includegraphics[trim=0 10 0 14,width=\columnwidth]{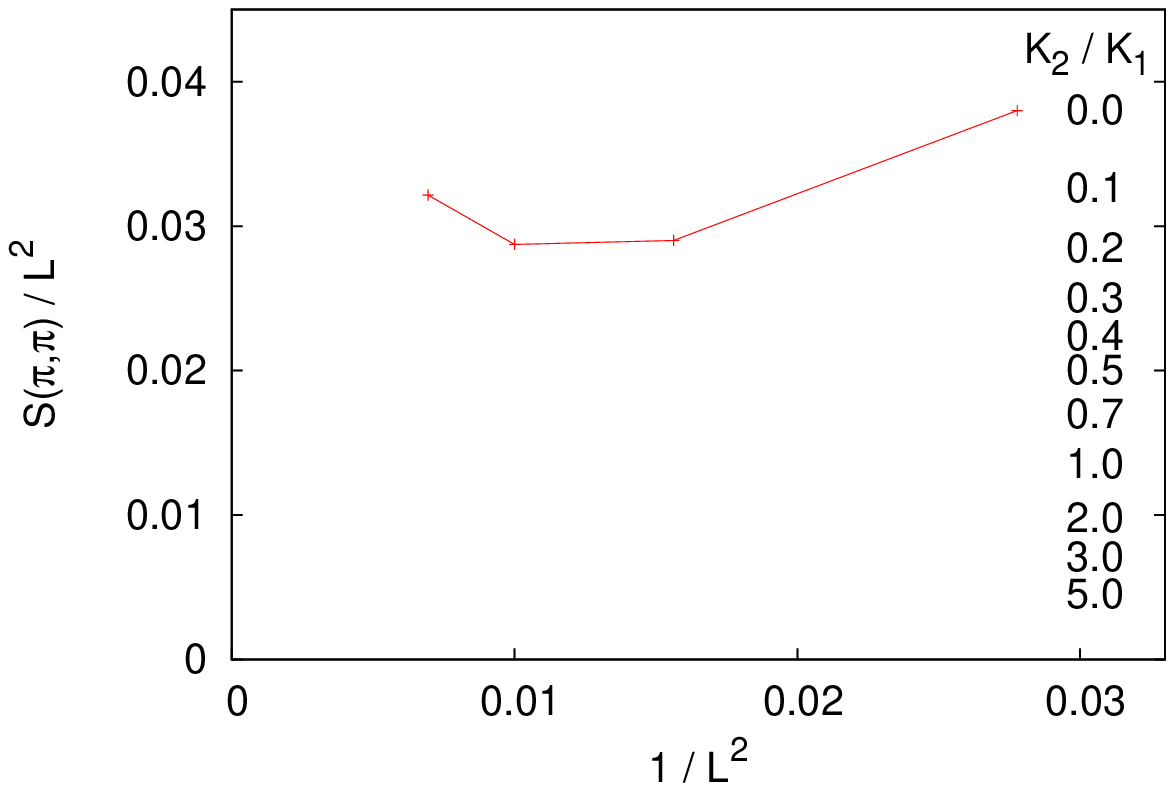}\\
  \includegraphics[trim=0 10 0 0,width=\columnwidth]{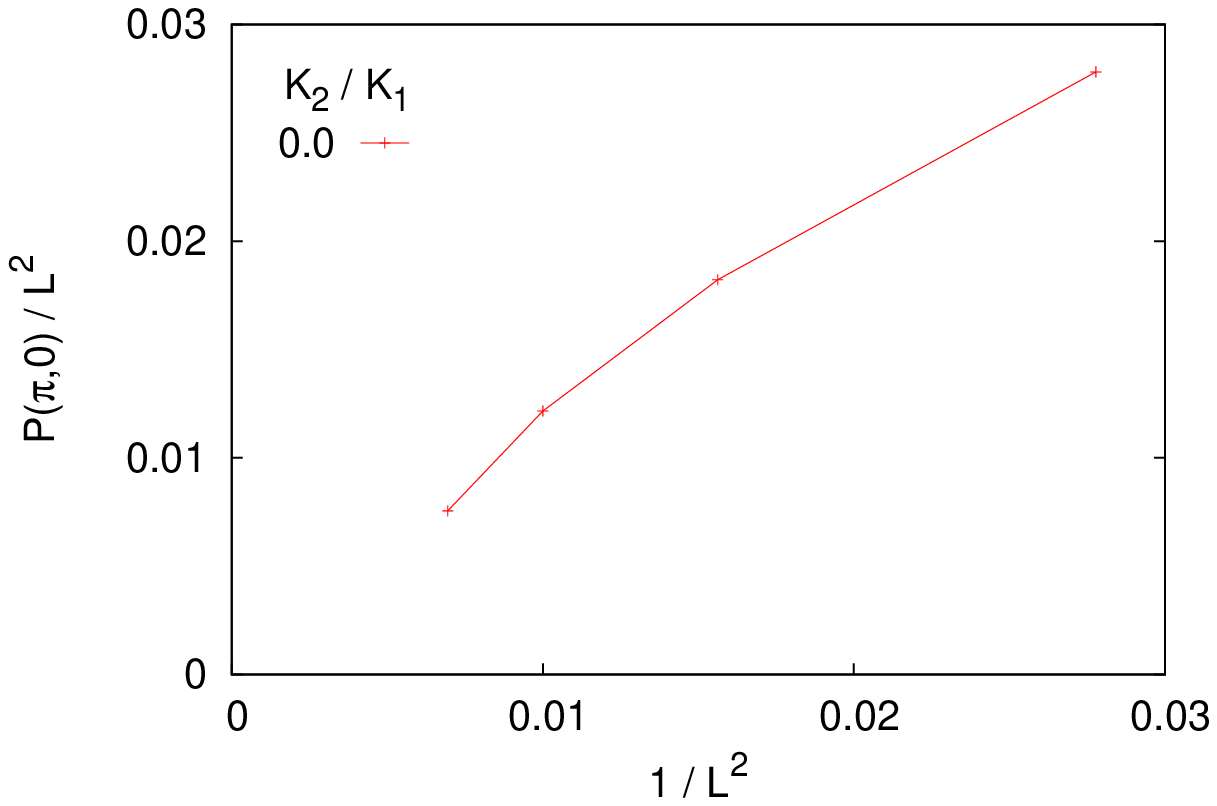}
  \caption{[Color online] Top: Finite size scaling of the staggered CDW order parameter $S(\pi,\pi)/L^2$ at selected $K_2$. Bottom: Finite size scaling of the bond-solid order parameter $P(\pi, 0)/L^2$ at selected $K_2$.}
  \label{fig:SF_L}
\end{figure}

Figure~\ref{fig:SF_crossover} shows the density structure factor $S(\pi,\pi)$ and the plaquette structure factor $P(\pi,0)$ plotted against $K_2$ for lattice sizes ranging from $L=6$ to $12$.  At small $K_2$, $S(\pi,\pi)$ increases strongly with $L$ while the size dependence becomes weaker at intermediate $K_2$.  This coincides with a strengthening size dependence of $P(\pi,0)$.  Figure~\ref{fig:SF_L} shows the finite size scaling of the respective order parameters.  For $K_2 < 0.4$, the CDW order is present while the plaquette order is absent.  For $K_2 > 0.4$, the situation is reversed.  Thus we conclude that the system undergoes a transition from the $(\pi, \pi)$ CDW at small $K_2$ to the striped $(\pi,0)$ [or $(0,\pi)$] plaquette order at intermediate $K_2$, where the latter is consistent with either a columnar VBS or a plaquette state.

To get a complete picture, we measure the structure factors over the entire Brillouin zone, paying attention also to long-wavelength behavior near the lines $q_x = 0$ and $q_y = 0$.  The CDW is an incompressible Mott insulator and is expected to have analytic $S(q_x, q_y)$.  This is indeed what we observe \cite{long_paper} and is in contrast with the compressible EBL behavior described below.  The Mott-like dependence of $S(q_x,q_y)$ at long wavelengths persists also after the $(\pi,\pi)$ CDW Bragg peak disappears for $K_2 \gtrsim 0.4$, thus indicating that the system is still not a liquid but a different solid (here bond-solid as determined by the corresponding order parameters).

For the plaquette structure factor $P(q_x, q_y)$, we find that $(\pi,0)$ and $(0,\pi)$ are the only sharp peaks.  We also measure bond energy structure factor (not shown) and again find only $(\pi,0)$ or $(0,\pi)$ order.  This is more consistent with the columnar VBS state, which is similar to the phase in the $J$-$K$ model\cite{Sandvik2002, Melko2004}.

\begin{figure}
  \centering
  \includegraphics[trim=15 20 10 50,width=\columnwidth]{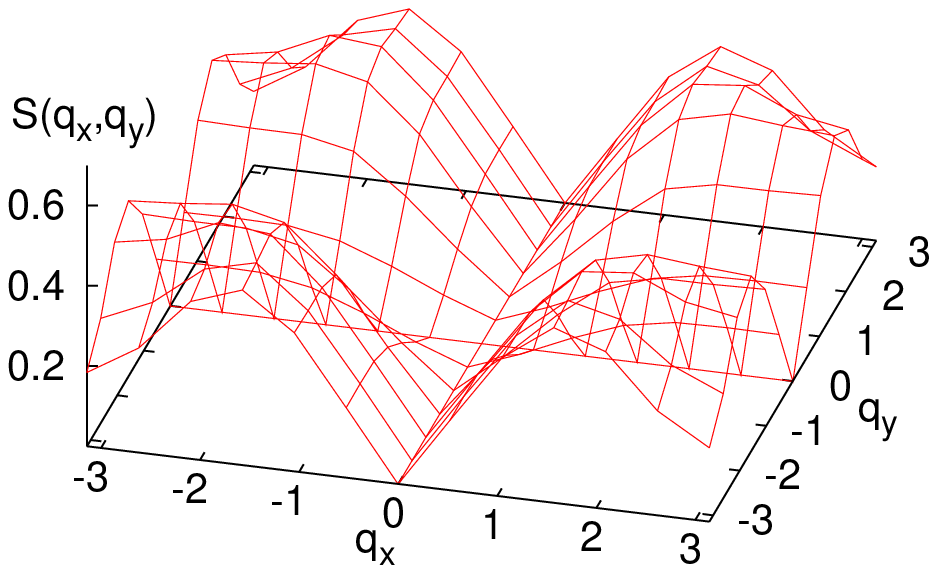}\\
  \includegraphics[trim=00 10 10 40,width=\columnwidth]{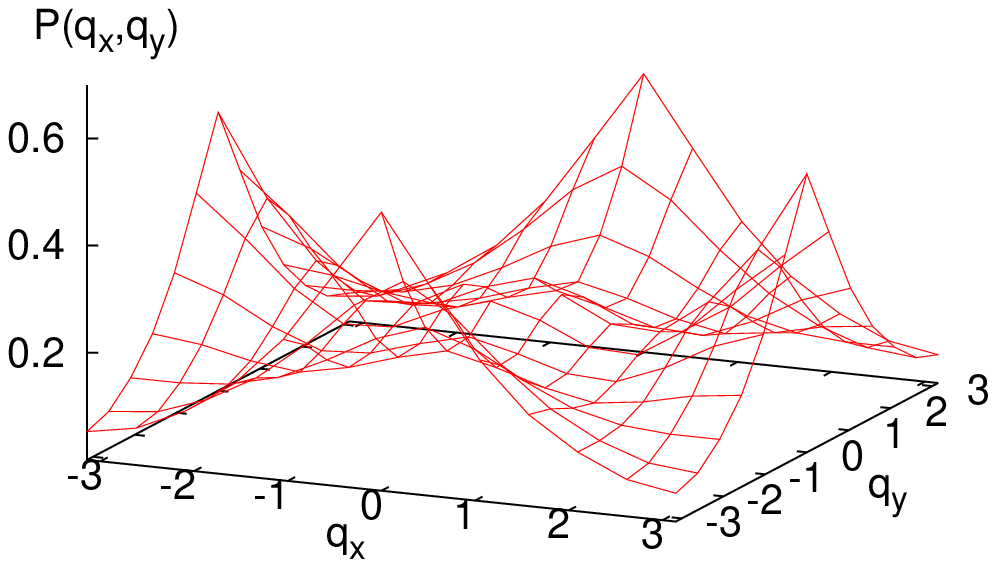}\\
  \includegraphics[trim=00 10 10 0,width=3in]{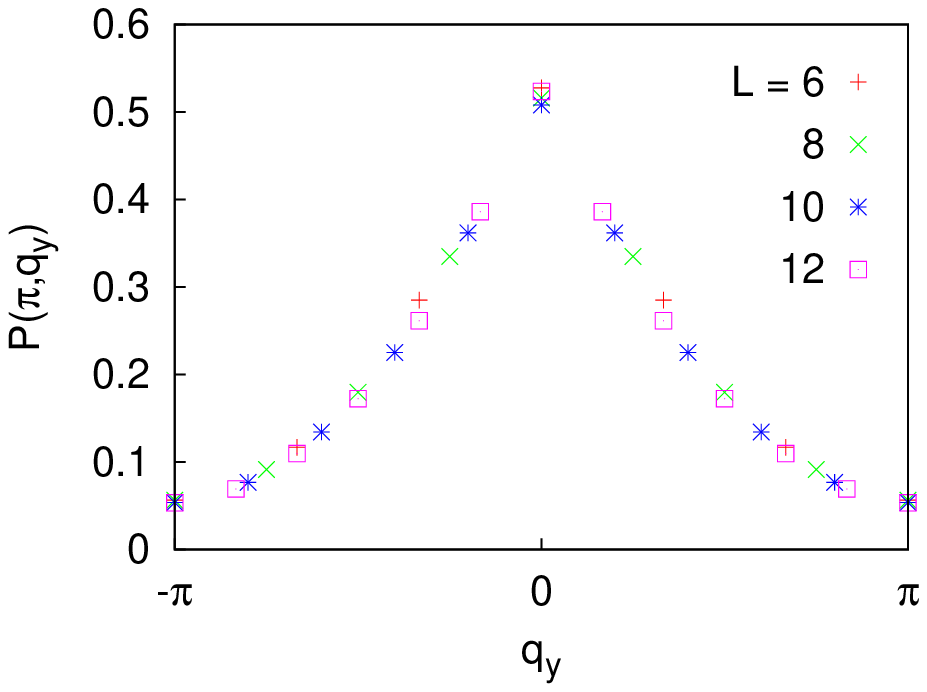}
  \caption{[Color online] Top: Density structure factor $S(q_x, q_y)$ for a 12$\times$12 lattice at $K_2=5$.  Middle: Plaquette structure factor $P(q_x, q_y)$ for the same system.  Bottom: Cut of the plaquette structure factor at $q_x=\pi$ for lattice sizes $L=6,8,10$ and $12$.}
  \label{fig:SF_12x12_k5}
\end{figure}

At larger $K_2 \gtrsim 4$, the plaquette structure factor is essentially insensitive to lattice size, and we conclude, within the limitation of our small lattice study, that the $(\pi,0)$ plaquette Bragg peak is absent.  This suggests that the bond-solid ordering exists only at intermediate $K_2$ and opens the possibility of the EBL phase for large $K_2$.

Our present evidence for the EBL comes from the absence of any charge or bond order.  Specifically, we show the structure factors $S(q_x,q_y)$ and $P(q_x,q_y)$ over the entire Brillouin zone.  Top panel in Fig.~\ref{fig:SF_12x12_k5} shows the density structure factor at $K_2=5$ and clear absence of any CDW ordering. Middle panel in Fig.~\ref{fig:SF_12x12_k5} shows the plaquette structure factor for the same system, which again does not show bond or plaquette ordering.  Despite the potential instability hinted by the $P(\pi,0)$ and $P(0,\pi)$ cusps, the nearly size independence of the plaquette structure factor illustrated along the cut $q_x=\pi$ shown in the bottom panel of Fig.~\ref{fig:SF_12x12_k5} gives us some confidence that there is no bond-solid order, so the EBL phase may indeed be realized in the large $K_2$ regime at half-filling.

Furthermore, we highlight the presence of the long wavelength EBL signature in the density structure factor near the lines $q_x=0$ and $q_y=0$.  The EBL theory \cite{Paramekanti2002} predicts a V-shaped singularity \cite{footnote} for small $q_x$ across a cut at a fixed $q_y$,
\begin{equation}
S_{\rm EBL} (q_x \to 0, q_y) = c(q_y) |q_x| ~.
\label{S_EBL}
\end{equation}
The slope $c(q_y)$ is non-zero except in the limit of small $q_y$ where $c(q_y) \sim |q_y|$.  The EBL behavior is qualitatively different from the incompressible non-singular behavior observed in the CDW and VBS regimes.  Our present data in the $K_2 \gtrsim 4$ regime shows large and healthy slopes $c(q_y)$ consistent with the EBL.  However, we caution that there is some weak downwards renormalization of the slopes upon increasing $L$ and so far we cannot rule out the possibility of the EBL behavior disappearing at much larger lattice sizes, which would then hint that there is bond-solid order with very small order parameter or some other instability.

We extend the search for the EBL phase to densities $n<1/2$.  Previous studies \cite{Rousseau2004, Melko2004, Rousseau2005} of the $J$-$K$ model have found that ring interactions induce strong tendencies to phase separation. For large $K$, uniform superfluid exists only for $n \gtrsim 0.4$, while the bosons phase separate at lower densities.  The more extended $K_2$ ring interactions can somewhat offset this tendency and stabilize a uniform liquid over a larger range of densities below half-filling.
Here we discuss our results for two values $K_2=0.5$ and $1.0$ on a 12$\times$12 lattice.   We consider the model with densities varying from 2 up to 5 bosons per row and column (corresponding to densities $\bar{n} = 1/6$ up to $5/12$).  Note that despite the condition of equal number of bosons in each row and column, the system can still phase-separate where bosons clump, e.g., along the $\hat{x}+\hat{y}$ direction in our samples.  (Of course, in a finite system, this condition frustrates somewhat the tendencies to phase separation, but we hope that already with our sizes we can observe the local energetic preferences.)
For 2 and 3 bosons per row at $K_2 = 0.5$ and 2 bosons per row at $K_2 = 1.0$, we indeed see such signatures of the phase separation in real space and also in momentum space, where the density structure factor develops strong peaks at the smallest wavevectors $2\pi/L$ \cite{long_paper}.  On the other hand, for the higher boson numbers per row, we do not see any signs of the phase separation or some order.  The density structure factor near $q_x=0$ or $q_y=0$ lines shows clear signatures of the EBL as discussed after Eq.~(\ref{S_EBL}) with robust and essentially size-independent slopes \cite{long_paper}, i.e., we do not see the weak downward renormalizations of the slopes that constitute one nagging worry about our EBL for large $K_2$ at half-filling discussed earlier.
Although our study is performed on small lattices only, it strongly suggests that the EBL phase is realized for densities $1/3 \leq \rho < 1/2$ for intermediate $K_2$ values, while the phase separation sets in at lower density.


We can rationalize our findings using the EBL theory \cite{Paramekanti2002,Xu2004, Xu2005, Nussinov2006, Xu2007}.  In a longer paper \cite{long_paper}, we will present an interesting perspective on the EBL using slave particle approach with two partons propagating one-dimensionally (1D) along one or the other lattice directions, which in particular allows application of quasi-1D thinking to this problem.  The EBL fixed point theory is characterized by a ``stiffness'' and the stability requires this stiffness to be sufficiently large.  The condition is particularly stringent at half-filling because of allowed Umklapp interactions.  When the system is unstable, it is driven to either CDW or plaquette solid depending on the sign of some effective interactions.  A crude scenario of what may be happening in our $K_1$-$K_2$ model is that for small to intermediate $K_2$ the system is unstable to Umklapps.  For small $K_2$, the effective nearest-neighbor repulsion (due to avoidance preferred by the $K_1$ ring exchanges) drives the system to the CDW state.  Increasing $K_2$ effectively introduces a competing second-neighbor repulsion and switches the instability to the plaquette solid (our measurements suggest that a columnar VBS is realized instead, but the two bond-solids are usually closely related).  Overall, increasing $K_2$ makes the bare stiffness large, as we observe from monitoring the density structure factor near the lines $q_x = 0$ or $q_y = 0$, and for large enough $K_2$ the EBL becomes stable even at half-filling.  When we step away from half-filling, the Umklapp terms are no longer allowed, and we can get a stable EBL already for intermediate $K_2$.


To conclude, our study of the $K_1$-$K_2$ model shows the transition from the CDW to VBS order at intermediate $K_2$ and tantalizingly suggests the EBL phase at large $K_2$.  Further studies of larger sizes and confrontation with the theory of this unusual phase will certainly be very interesting.  Even if the $K_1$-$K_2$ model has only solid phases, we have come much closer to the EBL and small additional terms should be able to further stabilize it.  The EBL phase is more robust away from half-filling and is realized in the $K_1$-$K_2$ model already for intermediate $K_2$.


We are thankful to M.~P.~A.~Fisher for inspiring discussions and encouragement throughout this work.  The research is supported by the NSF through grant DMR-0907145 and the A.~P.~Sloan Foundation.  We also acknowledge using the IT2 cluster at the Caltech CACR.


\end{document}